\documentclass[aps,floatfix,nofootinbib,amsmath,amssymb,amsfonts,notitlepage,superscriptaddress, twocolumn]{revtex4-2}

\usepackage[T1]{fontenc}
\usepackage[latin9]{inputenc}
\usepackage{lmodern} 
\usepackage{color}
\usepackage{bbold}
\usepackage{dsfont}
\usepackage{graphicx}
\usepackage{comment}
\usepackage[caption=false]{subfig}
\usepackage[colorlinks]{hyperref}
\usepackage[bottom]{footmisc}
\usepackage{enumerate}
\usepackage{float}
\usepackage{mathtools}
\usepackage{accents}

\usepackage{empheq}
\usepackage{tikz}
\usepackage[normalem]{ulem}
\DeclareMathAlphabet\mbc{OMS}{cmsy}{b}{n}
\usepackage{balance}

\begin{document}

\newcommand{\ks}[1]{{\textcolor{teal}{[KS: #1]}}}
\newcommand{\cj}[1]{{\textcolor{blue}{CJ: #1}}}
\newcommand{\ab}[1]{{\textcolor{green}{[AB: #1]}}}

\global\long\def\eqn#1{\begin{align}#1\end{align}}
\global\long\def\com#1{}
\global\long\def\vec#1{\overrightarrow{#1}}
\global\long\def\ket#1{\left|#1\right\rangle }
\global\long\def\bra#1{\left\langle #1\right|}
\global\long\def\bkt#1{\left(#1\right)}
\global\long\def\sbkt#1{\left[#1\right]}
\global\long\def\cbkt#1{\left\{#1\right\}}
\global\long\def\vbkt#1{\left|#1\right|}
\global\long\def\abs#1{\left\vert#1\right\vert}
\global\long\def\cev#1{\overleftarrow{#1}}
\global\long\def\der#1#2{\frac{{\mathrm{d}}#1}{{\mathrm{d}}#2}}
\global\long\def\pard#1#2{\frac{{\partial}#1}{{\partial}#2}}
\global\long\def\re{\mathrm{Re}}
\global\long\def\im{\mathrm{Im}}
\global\long\def\dd{\mathrm{d}}
\global\long\def\ddd{\mathcal{D}}
\global\long\def\hmb#1{\hat{\mathbf #1}}
\global\long\def\avg#1{\left\langle #1 \right\rangle}
\global\long\def\mr#1{\mathrm{#1}}
\global\long\def\mb#1{{\mathbf #1}}
\global\long\def\mc#1{\mathcal{#1}}
\global\long\def\tr{\mathrm{Tr}}
\global\long\def\dbar#1{\Bar{\Bar{#1}}}

\global\long\def\ubar#1{\underaccent{\bar}{#1}}
\global\long\def\ii{\mathrm{i}}

\global\long\def\nth{$n^{\mathrm{th}}$\,}
\global\long\def\mth{$m^{\mathrm{th}}$\,}
\global\long\def\non{\nonumber}

\global\long\def\ubar#1{\underaccent{\bar}{#1}}

\newcommand{\orange}[1]{{\color{orange} {#1}}}
\newcommand{\cyan}[1]{{\color{cyan} {#1}}}
\newcommand{\blue}[1]{{\color{blue} {#1}}}
\newcommand{\yellow}[1]{{\color{yellow} {#1}}}
\newcommand{\green}[1]{{\color{green} {#1}}}
\newcommand{\red}[1]{{\color{red} {#1}}}
\global\long\def\todo#1{\orange{{$\bigstar$ \cyan{\bf\sc #1}}$\bigstar$} }

\title{Decoherence of  spatial superpositions along stationary worldlines}

\author{Clemens Jakubec }
\email{clemens.jakubec@tuwien.ac.at}
\affiliation{Vienna Center for Quantum Science and Technology, Atominstitut, TU Wien, Stadionallee 2, 1020 Vienna, Austria}

\author{Aaron Bartleson}
\email{abartleson@arizona.edu}
\affiliation{Department of Physics, University of Arizona, Tucson, Arizona 85721, USA}
\affiliation{Wyant College of Optical Sciences, University of Arizona, Tucson, Arizona 85721, USA}

\author{Peter W. Milonni}
\email{peter_milonni@comcast.net}
\affiliation{Department of Physics and Astronomy, University of Rochester, Rochester, New York 14627, USA}

\author{Kanu Sinha}
\email{kanu@arizona.edu}
\affiliation{Department of Physics, University of Arizona, Tucson, Arizona 85721, USA}
\affiliation{Wyant College of Optical Sciences, University of Arizona, Tucson, Arizona 85721, USA}

\begin{abstract}
We analyze the decoherence of a particle's spatial superposition moving along a stationary worldline through the Minkowski vacuum. The particle is modeled via an internal degree of freedom that couples to a scalar field, and an external degree of freedom, i.e., its quantized center-of-mass motion around the stationary worldline. Assuming a separation of time scales between the particle's internal and external dynamics, we first obtain an effective red-shifted polarizability of the particle, characterizing the trajectory-dependent linear response of the internal oscillator to the field. We then derive a quantum Brownian motion master equation for the particle's center of mass, under the Born-Markov approximation, which describes its decoherence in the position basis, as well as Hamiltonian modifications corresponding to a dispersive potential. The resulting decoherence has two components:  (1) arising from a modified field spectrum observed by the particle; and (2) due to a differential time dilation over the particle's extended spatial wavefunction. For stationary trajectories, both contributions take an effectively thermal form. We evaluate the decoherence rates for two specific cases, hyperbolic and uniform circular motion.
\end{abstract}

\maketitle

\section{Introduction}
A particle undergoing uniformly accelerated linear motion through the Minkowski vacuum of a quantum field will perceive a thermal field environment. This effective thermality of the environment, also known as the Davies-Unruh effect \cite{Fulling_1973, Davies_1975, Unruh_1976}, raises the question: what is the backaction of such a thermal environment on the particle? It has been shown, for example, that the effective thermal field  can induce a range of remarkable effects such as momentum recoil \cite{Kempf_2021}, Casimir-Polder forces between atoms \cite{Passante_2014} and acceleration-induced transparency and a stimulated Unruh effect \cite{Kempf_2022}.
Recent works have also demonstrated that the backaction of such an effective field  leads to a momentum diffusion and drag on a polarizable particle, equivalent to that of a particle immersed in a thermal bath at the Davies-Unruh temperature $T_\mr{DU}=\hbar a/(2\pi k_\mr{B} c)$ \cite{PWM_KS2024}. Considering that the momentum diffusion of a polarizable particle in thermal radiation is  linked to decoherence of its quantized center-of-mass \cite{PWM_KS22, Jakubec_2025}, it is natural to ask: how does a spatial superposition decohere as it undergoes accelerated motion?

\begin{figure*}[t]
    \centering
    \includegraphics[width=0.9\textwidth]{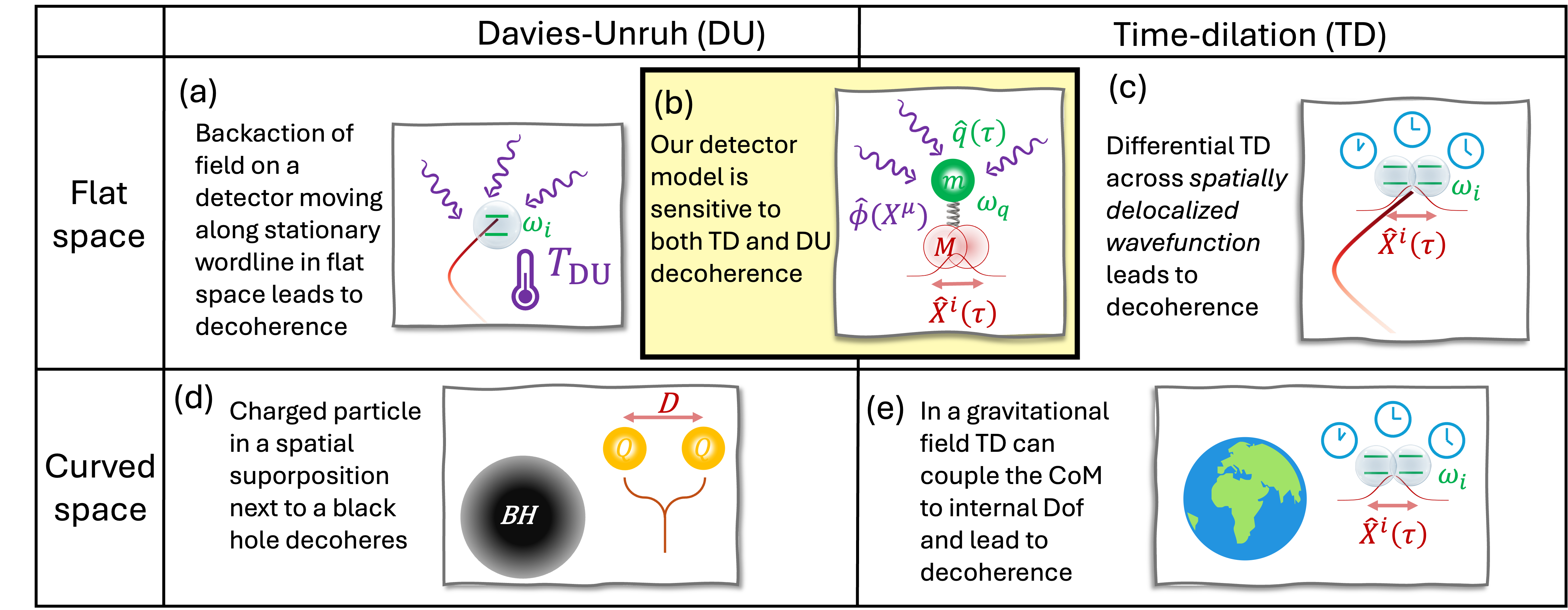}
    \caption{Summary of commonly studied decoherence mechanisms in flat spacetime as well as curved spacetime. (a): Unruh-DeWitt field detectors with a classical center of mass consisting of either a two-level system (with frequency $\omega_i$) or a harmonic oscillator have been studied in a plethora of situations. Along a hyperbolic worldline such a detector will see an effective temperature $T_\mr{DU}$. The thermalization and decoherence of such detectors has been studied in \cite{Parentani_1995,Lin_Hu_2006,Fewster_2016,Juarez-Aubry_2019,Foo_2020, Perche_2021, Foo_2025,Sudhir_2025,Stargen2026}. (b): Our model (see Fig. \ref{fig:schematic}) where a particle with an internal oscillator degree of freedom in a spatial superposition of $\hat{X}^i(\tau)$ is coupled to a field via its polarizability combines the decoherence mechanisms of (a) and (c). (c): Even in the absence of an external field, a detector's internal degree of freedom can couple to its quantized center of mass via time dilation. This effect has been described in \cite{Pikovski2015,Pikovski_2017,Paczos_2024}. (d): A charged particle without an internal degree of freedom, brought into a spatial superposition of size $D$ next to a black hole, will decohere due to the emission of photons across the black hole horizon \cite{DSW20222, DSW_Killing, Wei2024, Chen2024}. (e): Similar to (c), the time dilation in a gravitational field can couple a system's center of mass to its internal degree of freedom without a system-field coupling \cite{Pikovski2015,Pikovski_2017,Paczos_2024}. }
    \label{fig:table} 
\end{figure*}

 A variety of models have been considered  for investigating decoherence of particles moving non-inertially in flat spacetime, as well as in curved spacetime, as depicted in  Fig~\ref{fig:table}. In general, two different decoherence mechanisms arise: (1) the direct backaction of the modified field spectrum onto the system causes it to decohere, here referred to as \textit{Davies-Unruh decoherence}; and (2) the differential time dilation over the particle's extended wavefunction, which engenders which-path information.  We will refer to this as \textit{time-dilation decoherence}. 

Davies-Unruh decoherence has been studied in the context of accelerated motion through flat spacetime \cite{Parentani_1995,Lin_Hu_2006,Fewster_2016,Juarez-Aubry_2019,Foo_2020,Perche_2021,Foo_2025,Sudhir_2025,Stargen2026} and in connection with black holes, where quantum decoherence near black-hole event horizons offers a compelling window into the interplay between quantum information and general relativity. It is well known that after matter falls into a black hole, it emerges as thermal Hawking radiation, losing any coherence \cite{Unruh_Wald_2017}; striking recent results have demonstrated that the mere presence of an event horizon decoheres a spatial superposition nearby \cite{DSW20222, DSW_Killing, Wei2024, Chen2024}. Such decoherence of charged (massive) particles is attributed to the measurement of `which-path' information by the black-hole horizon via photons (gravitons). Thus, it has been conjectured that a particle in the vicinity of a horizon will experience decoherence both from the presence of the horizon as well as from the Hawking radiation emitted. 
Time-dilation decoherence, similarly, has been explored in the context of accelerated motion and gravitational fields \cite{Pikovski2015,Pikovski_2017,Paczos_2024}. Here it was shown that a system with internal degrees of freedom as well as a quantized center of mass will experience a system-bath coupling induced by the differential time dilation across the center-of-mass wavefunction, where the center of mass plays the role of the system and the internal degrees of freedom act as the bath. 

In this paper, we  analyze the open system dynamics of the quantized center-of-mass motion of a particle moving along a stationary worldline and interacting with a scalar field~\cite{Letaw_1981_stationary, Bunney_2024_stationary}. We show that the effective polarizability of the particle undergoing non-inertial motion experiences an effective redshift.  We derive a quantum master equation for the center of mass, and show that the interaction of the particle with its effective environment results in: (1) a  dispersive  potential, and (2) decoherence and dissipation. The decoherence seen by the particle results both from the modified spectrum of the field seen by the particle (Davies-Unruh decoherence), as well as the differential time dilation over the particle's wavefunction (time-dilation decoherence).

The remainder of this paper is organized as follows: In Section \ref{sec:model} we describe our model and derive the linearized interaction Hamiltonian between the quantized center-of-mass motion and its effective environment, encompassing the particle's internal degrees and the field. In Section \ref{sec:Open System Dynamics} we derive the quantum Brownian motion master equation and calculate the decoherence rates arising from  two distinct mechanisms: Davies-Unruh decoherence and time dilation. In Sections \ref{sec:hyperbolic_motion} and \ref{sec:circular_motion} we present calculations of the decoherence under uniform hyperbolic and circular motion, respectively. Finally, we discuss our results and give an outlook in Section \ref{sec:discussion_outlook}.

\section{Model}
\label{sec:model}

Let us consider a particle moving along a worldline \(x_0^\mu(\tau)\) in Minkowski space. The particle is composed of an internal oscillator degree of freedom \(\hat q\) (conjugate momentum \(\hat p\)) and an external quantized center-of-mass degree of freedom around the worldline $x^\mu_0(\tau)$, which we express in the local rest frame of the particle and denote by $ \hat X^i $ (conjugate momentum $ \hat P_i $) \footnote{For simplicity, and to avoid the pathologies that arise from writing a relativistic position operator (see, e.g., \cite{Hegerfeldt_1,Hegerfeldt_2}), we treat \(\hat X^i\) and \(\hat P_i\) as nonrelativistic observables. This amounts to expressing \(\hat X^i\) and \(\hat P_i\) in  local normal coordinates around \(x_0^\mu(\tau)\) and demanding that the particle's position wavefunction be confined to within a length scale  $\ell \sim  c^2/\sqrt{a_\mu a^\mu}$, where $a^\mu$ is the proper acceleration along $x_0^\mu(\tau)$ \cite{Perche_2022}.}. We will assume the particle's worldline $x_0^\mu(\tau)$ to be \textit{stationary}, such that the state of the field observed by the detector has a time-independent energy spectrum~\cite{Letaw_1981_stationary,Bunney_2024_stationary}.
Here Greek indices $(\mu, \nu, \mr{etc}.)$ are spacetime indices, whereas Latin indices $(i,j, \mr{etc.})$ are spatial indices.

\begin{figure}[t]
    \centering
    \includegraphics[width=0.9\linewidth]{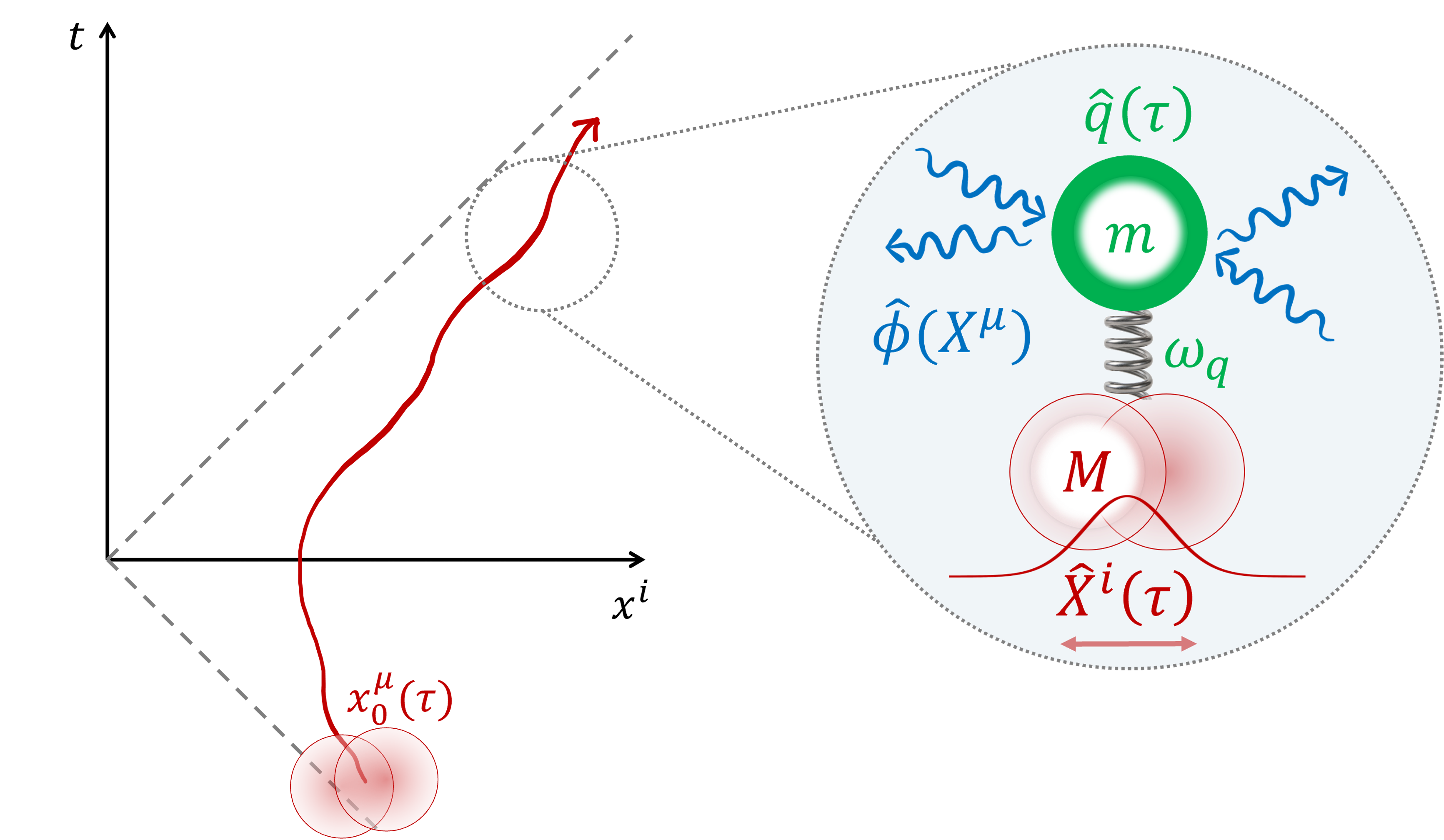}
    \caption{ Schematic representation of a  particle with mass $M$ moving along a worldline ${X}^\mu_0 (\tau)$ through Minkowski spacetime. The particle is prepared in a spatial superposition, with  $\hat X^i  (\tau)$ as its quantized center-of-mass position. The particle has an internal oscillator degree of freedom with position $ \hat q (\tau )$, mass $m$, and frequency $\omega_q$, and couples to a scalar field $ \hat \phi (\hat X^\mu (\tau ))$.}
    \label{fig:schematic}
\end{figure}

The internal oscillator represents a charge-like degree of freedom and is coupled to a 3+1 dimensional scalar field at the position of the particle (see Fig. \ref{fig:schematic}), which is given in the local rest frame of the particle as:
\begin{align}
\hat{\phi}(X^\mu)=\sum_\mb{k}\sbkt{\hat g_\mb{k}\phi_\mb{k}(X^\mu)+\hat g_\mb{k}^\dagger\phi^\ast_\mb{k}(X^\mu)  }.
\end{align}
Here $\hat g^\dagger_\mb{k}$ and $\hat g_\mb{k}$ are the creation and annihilation operators associated with the mode functions $\phi_\mb{k}(X^\mu)$ and $\phi^*_\mb{k}(X^\mu)$, following the canonical commutation relation $ \sbkt{\hat g_\mb{k},\hat g_{\mb{k}'}^\dagger} = \delta_{\mb{k},\mb{k}'}$. 

In the local rest frame of the particle, the total Hamiltonian is
$\hat H=\hat{H}_\mr S+\hat{H}_\mr F$,
with the field Hamiltonian $\hat H_\mr F$ given by
\begin{align}
\hat H_\mr F=\sum_\mb{k}\hbar \Omega_\mb{k} \hat g^\dagger_\mb{k}\hat g_\mb{k}.
\end{align}

The system Hamiltonian in Schr\"odinger picture $\hat{H}_\mr S$ includes the rest energy of the particle, the kinetic energy of the center of mass as well as the interaction of the particle with the field  (see Appendix~\ref{App:Lag+Ham} for details):
\begin{align}
\hat{H}_\mr S= \sqrt{-g_{00}(\hat X^i)\bkt{\hat E_\mr{rest}^2+c^2\hat P_i\hat P^i}},
\end{align}
where $g_{00}(\hat X^i)$ is the $00$-component of the locally static metric $g_{\mu\nu}$ and describes the gravitational time dilation experienced by the internal oscillator degree of freedom, with $\tau$ as the proper time of the moving particle. The rest energy of the system is
\begin{align} 
\label{eq:Erest}
\hat E_\mr{rest}=\mathbb{1} \ Mc^2+\frac{1}{2m}\sbkt{\hat p-e\hat\phi(\hat X^i)}^2+\frac12 m\omega_q^2\hat q^2, 
\end{align}
with $M$ the rest mass and $\omega_q$ the internal oscillator frequency seen by a comoving observer. The internal oscillator couples to the field $\hat \phi(\hat X^i)$ at the position of the particle, with a strength $e$.
Under the assumption that  $\Delta P_i/(Mc) \ll 1$ in the particle's rest frame, i.e.  the quantized center-of-mass motion of the particle in the proper frame is nonrelativistic, we can simplify the system Hamiltonian to
\eqn{ \label{eq:Hs_NR}
\hat{H}_\mr{S}&\approx\sqrt{-g_{00}(\hat X^i)}\bkt{\hat E_\mr{rest}+ \frac{\hat P_i\hat P^i}{2M}}.
}
We now make a separation of time scales assumption, considering the internal oscillator dynamics to be much faster than the quantized center-of-mass motion. We first determine  the coupled dynamics of the internal oscillator and the  field by solving  their coupled Heisenberg equations of motion \cite{PWM_KS2024} (see Appendix~\ref{App:intfield}):
\begin{align}
    \ddot{\hat q}-g_{00}\omega_q^2\hat q=-\sqrt{-g_{00}}\frac{e}{m}\dot{\hat \phi}_\mr{tot}(\hat X^i,\tau).
\end{align}
We solve for the dynamics of the internal oscillator, as driven by the field,  assuming that the internal degree of freedom has a sufficiently fast equilibration time such that its dynamics is pinned to that of the field. Thus in stationary state we can describe the oscillator amplitude in terms of an effective induced dipole as  (see Appendix~\ref{App:polarizability} for details):
\begin{align}
\label{eq:dipole_spectral_decomp}
\hat d_\mr{st}(\hat X^i)\equiv& e \hat q _\mr{st}\non\\
=& -\frac{\ii}{2 \pi}\sqrt{-g_{00}(\hat X^i)}\int_{-\infty}^\infty \dd\omega \, \omega\alpha(\omega)\underaccent{\bar}{\hat \phi}(\hat X^i, \omega),
\end{align}
with a red-shifted linear response: 
\begin{align}\label{eq:redshift_polarizability}
    \alpha(\omega)\approx \alpha_0(\omega)\sbkt{1 - \eta(\omega) a_i \hat X^i /c^2}
\end{align}
and $\underaccent{\bar}{\hat \phi}(\hat X^i, \omega)$ defined via the spectral decomposition:
\begin{align}
    \hat{\phi}(\hat X^i)=\frac{1}{2\pi}\int_{-\infty}^\infty d\omega \underaccent{\bar}{\hat \phi}(\hat X^i, \omega).
\end{align}
Such an effective \textit{red-shifted polarizability}, which captures the linear response of a particle undergoing non-inertial motion, can be generalized to other particles (e.g., electrons).

Here we have neglected second and higher-order terms in $ \hat X ^i $, assuming that the quantized center-of-mass is sufficiently localized such that $ \Delta X ^i \ll c^2/a$.
The red-shifted polarizability stems from the fact that the characteristic frequencies experience a time dilation due to the spread of the particle's wavefunction around $x_0^\mu(\tau)$. Here $\alpha_0(\omega)$ is the bare polarizability in the rest frame of the particle and $\eta(\omega)$ is a redshift-induced correction given by: 
\begin{align}
    \alpha_0(\omega)=&\frac{e^2/m}{-\omega^2+2\ii\beta\omega^3+\omega_q^2}, \label{eq:alpha_0}\\ 
    \eta(\omega)=&\frac{2\ii\beta\omega^3+2\omega_q^2}{-\omega^2+2\ii\beta\omega^3+\omega_q^2}, \label{eq:eta}
\end{align}
where $\beta=e^2/(8\pi m c^3 \epsilon_0)$.\footnote{While we are not dealing with an electromagnetic (EM) field, the factor of $ \epsilon_0 $ arises from the choice of normalization of the field mode functions $ \cbkt{\phi_\mb{k},\phi_\mb{k}^\ast}$ such that $ \abs{\phi_\mb{k}}^2 = \frac{\hbar }{2 \epsilon_0\Omega_\mb{k}  V}$. With such normalization, one can establish a correspondence with the EM field by identifying the scalar field $ \hat \phi (X^\mu)$ as a component of the EM vector potential~\cite{Sinha2015, Sinha21}.} We point out that $\alpha_0(\omega)$ is technically not analytic in the upper half of the complex plane, due to the appearance of the $\omega^3$ term in the denominator. However, since $\beta$ is small compared to $1/\omega_q$, the contributions from this pole can generally be neglected. This allows us to simplify the interaction Hamiltonian  into a form similar to the electric-dipole form of the Hamiltonian (see Appendix~\ref{App:intfield} for a derivation).

Having solved for the coupled  dynamics of the internal oscillator and the field, and assuming a separation of time scales, we now turn to the slow quantized center-of-mass motion.  The quantized center-of-mass Hamiltonian consists of a kinetic center-of-mass contribution and an interaction term:
\begin{align}\label{eq:hs}
    \hat H_\mr{M}
    &\equiv \sqrt{-g_{00}(\hat X^i)}\bkt{Mc^2+\frac{\hat P_i\hat P^i}{2M}}
    \non\\&-\frac{1}{4} \cbkt{\dot{\hat{d}}_\mr{st}(\hat X^i),\hat \phi(\hat X^i)},
\end{align}
where we  use the steady-state dynamics of the internal oscillator to determine its response to the field (Eq.~\eqref{eq:dipole_spectral_decomp}). Note that the coupling term between dipole and field is symmetrized to ensure the hermiticity of \(\hat H_\mr{M}\).

We assume that the extent of the spatial wavefunction of our particle is small enough such that we can expand around the classical center-of-mass worldline of the particle, which now acts as a reference worldline. To make this expansion we write the metric in local normal coordinates \cite{Poisson2011}: 
\begin{align}    
g_{00}(\hat X^i) &\approx -\bkt{\mathbb{1}+\frac{a_i\bkt{\tau}}{c^2}\hat X^i}^2 \label{eq:FW_metric_00}-R_{0i0j}\bkt{\tau}\hat X^i\hat X^j,\\
g_{0i}(\hat X^i) &\approx -\frac23 R_{0jik} \hat X^j \hat X^k, \\
g_{ij}(\hat X^i) &\approx \delta_{ij}-\frac{1}{3}R_{ikjl}\bkt{\tau}\hat X^k \hat X^l,
\end{align}
with $a^\mu\bkt{\tau}$ the proper acceleration at $x_0^\mu(\tau)$ and $R_{\mu\nu\sigma\rho}\bkt{\tau}$ the curvature tensor evaluated on the particle worldline. We have ignored third and higher order terms in $\hat X^i$ in the above expansion. For an observer on a stationary worldline in Minkowski space the curvature tensor $R_{\mu\nu\sigma\rho}=0$. This allows us to study the quantum dynamics of the center of mass by evaluating field quantities on the prescribed classical worldline $x_0^\mu(\tau)$ only. Expanding the metric as well as the field to first order in $\hat X^i$ and neglecting terms proportional to $\hat P^i/(Mc)$ leaves us with the system Hamiltonian: 
\begin{align}
\label{Eq:ham2}
    \hat H_\mr M=& \mathbb{1} \ Mc^2-\frac{1}{4}\cbkt{\dot{\hat d}_{0, \mr{st}},\hat \phi}+\frac{\hat P_i\hat P^i}{2M}+Ma_i\hat X^i\non\\
    &-\frac12 \hat X^i \cbkt{\dot{\hat d}_{0, \mr{st}}, \partial_i\hat \phi}+\frac{a_i\hat X^i}{4c^2}\cbkt{\dot{\hat d}_{1, \mr{st}}-\dot{\hat d}_{0, \mr{st}}, \hat \phi}
\end{align}
where we use the shorthand notation $\hat \phi \coloneqq \hat \phi( X^i = 0)$ and define induced dipole moments in steady state:
\begin{align}
\label{eq:d0}
    \hat d_{0, \mr{st}}=&-\frac{\ii}{2\pi}\int_{-\infty}^\infty \dd\omega \, \omega\alpha_0(\omega)\underaccent{\bar}{\hat \phi}(\omega),\\
    \label{eq:d1}
    \hat d_{1, \mr{st}}=&-\frac{\ii}{2\pi}\int_{-\infty}^\infty \dd\omega \, \omega\alpha_0(\omega)\eta(\omega)\underaccent{\bar}{\hat \phi}(\omega).
\end{align}

Having written the Hamiltonian in Eq.~\eqref{Eq:ham2},  we can delineate  two different mechanisms that give rise to an interaction of the quantized center of mass with its  environment. Moving to the interaction picture (indicated via an explicit $\tau$ dependence of the observables),  we separate the total interaction Hamiltonian into two components: 
\begin{enumerate}
    \item The particle's quantized center-of-mass interacts with the field via the internal oscillator, as given by the following interaction Hamiltonian:
         \begin{align}
            \hat H_\mr{int}^\mr{DU}(\tau)\equiv-\frac12 \hat X^i(\tau) \cbkt{\dot{\hat d}_{0, \mr{st}}(\tau),\partial_i\hat \phi(\tau)}
            \label{eq:H_int_UD}
        \end{align}
    The  trajectory-dependent vacuum  state observed by the particle exerts a backaction  leading to center-of-mass decoherence.
    \item The differential time dilation across the center-of-mass wavefunction, described by the redshift-factor $g_{00}$, gives rise to the interaction Hamiltonian: 
    \begin{align}
        \hat H_\mr{int}^\mr{TD}(\tau)
        \equiv \frac{a_i\hat X^i(\tau)}{4c^2}\cbkt{\dot{\hat d}_{1, \mr{st}}(\tau)-\dot{\hat d}_{0, \mr{st}}(\tau), \hat \phi(\tau)}
        \label{eq:H_int_TD}
    \end{align}
    These interaction terms stem from the red-shifted response function $\alpha(\omega)$ of the particle. Although the nature of time-dilation induced coupling of the center of mass to a bath is akin to that in \cite{Pikovski2015}, we assume in our model that the internal degree of freedom has a sufficiently fast equilibration time such that its dynamics reaches a steady state and is thus pinned to that of the field. It follows that all terms in $\hat H_\mr{int}^\mr{TD}$ effectively amount to an interaction between the center of mass and the field.\footnote{The equilibration time of an Unruh-DeWitt detector has been studied in \cite{Stargen2026}. Here we assume that $\tau \gg \tau_\mr{thermal}$, such that transient field dynamics can be ignored.}
    
\end{enumerate}

\section{Open System Dynamics}
\label{sec:Open System Dynamics}
\subsection{Quantum Brownian Motion Master Equation}
In order to model the center-of-mass decoherence of our particle, we employ the Quantum Brownian Motion (QBM) master equation, which describes the reduced dynamics  of the quantized center-of-mass density matrix $\rho$.  We expand the interaction Hamiltonian in the interaction picture (indicated by an explicit time dependence) in small $\hat X^i$: 
\begin{align}
    \hat H_\mr{int}(\tau)&=\hat H_\mr{int}^0(\tau)+\partial_i\hat H_\mr{int}(\tau)\hat X^i+\mathcal{O}( (\hat X^i)^2)\non\\
    &=\hat H_\mr{int}^0(\tau) +\hat B(\tau)\hat X^i+\mathcal{O}((\hat X^i)^2),
\end{align}
with the bath operator   defined as:
\begin{align}
    \hat B_i(\tau)=\partial_i\hat H_\mr{int}(\tau)
\end{align}
Following our split of the interaction Hamiltonian into time-dilation and Davies-Unruh terms, we can do the same for the bath operator and define: 
\begin{align}
\label{eq:bath_operator_split}
    \hat{B}_i(\tau)=\hat{B}^\mr{DU}_i(\tau) + \hat{B}^\mr{TD}_i(\tau).
\end{align}
Here the time-dilation and Davies-Unruh bath operators are defined, respectively, as: 
\eqn{\label{eq:Bdu}
\hat{B}^\mr{DU}_i(\tau) &= -\frac12 \cbkt{\dot{\hat d}_{0, \mr{st}}(\tau),\partial_i\hat \phi(\tau)}; \\
\label{eq:Btd}
\hat{B}^\mr{TD}_i(\tau) &= \frac{a_i}{4c^2} \cbkt{\dot{\hat d}_{1,\mr{st}}(\tau)-\dot{\hat d}_{0,\mr{st}}(\tau), \hat \phi(\tau)}.
}

The QBM master equation resulting from the above system-bath interaction is thus~\cite{BPBook}: 
\begin{align}
\label{eq:QBM_master_equation}
\dot{\hat\rho}=&-\Lambda_{ij}\sbkt{\hat X^{i},\sbkt{\hat X^{j},\hat \rho}} -\frac\ii\hbar \Gamma_{ij}\sbkt{\hat X^{i},\cbkt{\hat P^{j},\hat \rho}} \non \\
&+ \frac\ii\hbar C^{(1)}_i\sbkt{\hat X^i,\hat\rho} - \frac\ii\hbar C^{(2)}_{ij}\sbkt{\hat X^{i}, \cbkt{\hat X^{j},\hat \rho}}
\end{align}
with $i,j$ corresponding to the three spatial dimensions in the rest frame of the particle; 
repeated indices are summed over. 

The derivation of the above QBM master equation relies on the validity of the Born-Markov approximation. This approximation applies for our model of a polarizable particle interacting with a field along a stationary worldline under two specific assumptions. Firstly, it was shown in \cite{Juarez-Aubry_2019} that an Unruh-DeWitt detector interacting with a scalar field moving on a stationary worldline reaches a steady state in the limit of $\tau\rightarrow\infty$. This, in addition to the assumption that the interaction between the particle and the field is sufficiently weak, allows us to make the Born-Markov approximation following the arguments in \cite{Lin_Hu_2006}.

In order, the terms in the above master equation represent: 
\begin{enumerate}
    \item The diagonal terms  $\Lambda_{ii}$ correspond to the decoherence in the position basis:
\begin{align}
\label{eq:deco_coeff}
\Lambda_{ij}=\frac{1}{2\hbar^2}\int_0^\infty \dd\tau' \avg{\cbkt{\hat B_i(\tau),\hat B_j(\tau-\tau')}}.
\end{align}
    \item The dissipation of the center-of-mass energy into the environment is
\eqn{
\Gamma_{ij}\equiv \frac{\ii}{2M\hbar}\int_0^\infty \dd\tau' \tau' \avg{\sbkt{\hat B_i(\tau),\hat B_j(\tau-\tau')}}.
}
    The decoherence rate $\Lambda_{ii}$ is related to $\Gamma_{ii}$ by the fluctuation-dissipation theorem: $2M\Gamma_{ii} k_\mr{B} T  = \hbar^2\Lambda_{ii}$.
    \item The system Hamiltonian is modified by the terms:
    \eqn{\label{eq:c1}
    C^{(1)}_i\equiv& \avg{ \hat B_i (\tau )},  \text{ and}\\ 
    \label{eq:c2}
    C^{(2)}_{ij}\equiv& \frac{\ii}{2\hbar}\int_0^\infty \dd\tau' \avg{\sbkt{\hat B_i(\tau),\hat B_j(\tau-\tau')}}.}
    Remarkably, these contributions are akin to the first derivative of the first order (in polarizability) Casimir-Polder potential ($C_i^{(1)} \sim \partial_i \mc{U}_\mr{CP}^{(1)}$), and second derivative of the second-order  Casimir-Polder potential ($C_{ij}^{(2)}\sim \partial_i \mc{U}_\mr{CP}^{(2)}\partial_j$) for a polarizable particle near a medium, as defined in \cite{Jakubec_2025}.  We note that while there is no actual medium or boundary present in the current model,  there is a nevertheless a \textit{dispersive potential or force } seen by the particle's quantized center of mass. 
\end{enumerate}


We will now examine the decoherence (\(\Lambda_{ii}\)) and dispersion potential ($ C_i^{(1)}$ and $ C_{ij}^{(2)}$) on the particle's center of mass.

\subsection{Decoherence}
Using the definition of the bath operators from the previous section and the interaction Hamiltonians Eqs.~\eqref{eq:H_int_TD} and \eqref{eq:H_int_UD}, we can now derive an expression for the total decoherence coefficient \eqn{\Lambda=\Lambda^\mr{DU}+\Lambda^\mr{TD}.}
Here  $\Lambda^\mr{DU}$ represents the `Davies-Unruh decoherence', arising from $\hat{H}_\mathrm{int}^\mr{DU}$, i.e., the backaction of the field on the particle,  and   $\Lambda^\mr{TD}$ corresponds to the `time dilation decoherence' due to $\hat{H}_\mathrm{int}^\mr{TD}$, i.e., the interaction between the field and the system from time dilation. Since the interaction Hamiltonian is quadratic in the field, the decoherence coefficients thus amounts to calculating fourth-order field correlators along the worldline of the particle. We make the following simplifying assumptions: (i) we assume that the field state observed by a particle moving along a stationary worldline is Gaussian, meaning that we can apply Wick's theorem to write the four-point correlators in terms of two-point correlators along the worldline of the particle. This also implies that the field and its spatial derivative are statistically independent, i.e. that $\langle\hat\phi(\tau)\partial_i\hat\phi (\tau-\tau')\rangle=0$~\cite{PWM_KS22}; and, (ii) we assume that different spatial directions of the center-of-mass are uncorrelated. This implies that $\langle\partial_j\hat\phi(\tau)\partial_i\hat\phi (\tau-\tau')\rangle=0$ for $i\neq j$, indicating that decoherence along different spatial directions occurs independently of one another. It follows from these assumptions that we can write the four-point correlators in terms of  $\langle\hat\phi(\tau)\hat\phi (\tau-\tau')\rangle$ and $\langle\partial_i\hat\phi(\tau)\partial_i\hat\phi (\tau-\tau')\rangle$ only.

\subsubsection{`Davies-Unruh' decoherence}

Armed with these assumptions, we can calculate the decoherence coefficient using Eq. (\ref{eq:deco_coeff}) and the expression for the Davies-Unruh bath operator from  Eq. (\ref{eq:H_int_UD}), as follows: 
\begin{align}
&\Lambda^\mr{DU}_{ij}=\frac{1}{2\hbar^2}\int_0^\infty \dd\tau'\left\langle\cbkt{\hat B_i^\mr{DU}(\tau),\hat B_j^\mr{DU}(\tau-\tau')}\right\rangle\non\\
=&\frac{1}{2\hbar^2}\int_0^\infty \dd\tau'\left\langle \dot{\hat d}(\tau)\dot{\hat d}(\tau-\tau')\right\rangle\left\langle \partial_i\hat \phi(\tau)\partial_i\hat \phi(\tau-\tau')\right\rangle\non\\
+&\frac{1}{2\hbar^2}\int_0^\infty \dd\tau'\left\langle \dot{\hat d}(\tau-\tau')\dot{\hat d}(\tau)\right\rangle\left\langle \partial_i\hat \phi(\tau-\tau')\partial_i\hat \phi(\tau)\right\rangle,
\end{align}
if $i=j$ and zero otherwise and we have used the assumptions stated earlier. To proceed we note that the correlator of the spatial derivatives can be turned into the correlator of the time derivative:
\begin{align}
\left\langle\partial_i\hat \phi(X)\partial_i\hat \phi(X')\right\rangle=\frac{1}{3c^2}\left\langle\partial_\tau\hat \phi(X)\partial_\tau\hat \phi(X')\right\rangle.
\end{align}
Since we are working in the local rest frame of the particle, the coordinate time derivative equals the proper time derivative. Spectrally decomposing the field  and evaluating the $\tau$ integral, we are left with (see Appendix~\ref{App:decoherence_coefficients} for details):
\eqn{
\label{eq:DU_deco}
\Lambda^\mr{DU}=\frac{1}{6\pi\hbar^2c^2}\int_0^\infty \dd\omega \, \omega^6 |\alpha_0(\omega)|^2 D^+(\omega)D^-(\omega).
}
Here $D^+(\omega)$ and $D^-(\omega)$ are the Fourier transforms of the forward and backward processes respectively, defined as: 
\eqn{
    D^+(\omega)=\int_{-\infty}^{\infty}\dd\tau'\avg{\hat{\phi}(\tau+\tau')\hat \phi(\tau)}e^{-i\omega\tau'},\non\\
    D^-(\omega)=\int_{-\infty}^{\infty}\dd\tau'\avg{\hat{\phi}(\tau-\tau')\hat \phi(\tau)}e^{-i\omega\tau'}.
    \label{eq:FT_correlators}
}
They satisfy the relation $D^+(-\omega)=D^-(\omega)$.

\subsubsection{Davies-Unruh Decoherence from Detailed Balance}
Remarkably, there is an analogy between the decoherence rate \(\Lambda^\mr{DU}\) derived above and the decoherence rate of an inertial detector coupled to an effective thermal bath, which suggests an alternative derivation. We start with the decoherence coefficient of an inertial, polarizable particle interacting with a thermal field \cite{PWM_KS22,Jakubec_2025}:
\eqn{
\label{eq:thermal_deco}
\Lambda_\mr{th}=\frac{1}{3 (2\pi)^3\epsilon_0^2 c^8} \int_0^\infty \dd \omega \, \omega^8 |\alpha_0(\omega)|^2 \bkt{n(\omega) + 1} n(\omega).
}
Here $n(\omega)=\sbkt{\text{exp}{(\hbar\omega/k_\mr{B}T)}-1}^{-1}$ is the Bose-Einstein distribution. In \cite{Juarez-Aubry_2019} it was shown that for any observer moving along a stationary worldline, the field state seen by the observer can be described by a temperature $T$ given by the detailed-balance relation: 
\eqn{
\label{eq:detailed_balance_stationary}
\frac{D^-(\omega)}{D^+(\omega)}=e^{ \hbar \omega/(k_\mr{B} T)}.
}
 One can thus derive the Davies-Unruh decoherence rate by starting from the expression for the decoherence in thermal radiation of a polarizable particle at rest in Minkowski spacetime given by Eq. (\ref{eq:thermal_deco}). Under the assumption that the processes in the environment are in detailed balance [Eq.(\ref{eq:detailed_balance_stationary})], we can write the the modified spectrum along a stationary worldline in the form of an effective Bose-Einstein distribution with a frequency-dependent temperature as \cite{Juarez-Aubry_2019, Good2020}: 
\eqn{
n(\omega)=\frac{1}{e^{\hbar \omega/(k_B T(\omega))}-1}
=\frac{D^+(\omega)}{D^-(\omega)-D^+(\omega)}, 
}
which immediately allows us to see that: 
\begin{align}
n(\omega)\bkt{n(\omega)+1}&=\frac{D^+(\omega)D^-(\omega)}{\bkt{D^+(\omega)-D^-(\omega)}^2}\non\\
&=\frac{(2\pi \epsilon_0 c^3)^2}{\hbar^2\omega^2}D^+(\omega)D^-(\omega), 
\end{align}
In the last step we have used the expression for the density of states in Minkowski spacetime. 

Since the Fourier transform of the pullback of the Minkowski spacetime field correlator onto any stationary worldline always satisfies a detailed balance relation, we can associate a temperature with the corresponding field state.  For instance, in the case of hyperbolic motion, this temperature would be equivalent to the well known Davies-Unruh temperature. Using Eq. (39) in Eq. (36), we recover Eq.(34). This tells us that the decoherence coefficient of our system can always be expressed in the form of Eq.(36) with an appropriate frequency dependent temperature $T(\omega)$, as long as the processes in the environment are in detailed balance.

\subsubsection{Decoherence from time dilation}

We similarly derive the contribution to the total decoherence arising from the  time-dilation interaction Hamiltonian in Eq.~\eqref{eq:H_int_TD}. As mentioned above, this interaction arises from the differential time-dilation over the extent of the spatial wavefunction of the system. We will thus refer to the associated decoherence effect as time-dilation decoherence, $\Lambda^\mr{TD}$. To calculate $\Lambda^\mr{TD}$, we first consider the time-dilation bath operator in Eq.~\eqref{eq:Btd}, wherein both terms depend on the dipole and are thus linear in the polarizability. We can immediately infer the resulting decoherence coefficient $\Lambda_{ij}^\mr{TD}$, by looking at the similarities between the second and third term in Eq. \eqref{eq:Btd} and Eq. \eqref{eq:Bdu}. 
In analogy to Eq. \eqref{eq:DU_deco}, the decoherence coefficient is thus: 
\begin{align}
    &\Lambda_{ij}^\mr{TD}=\non\\
    &\frac{a_ia_j}{6\pi\hbar^2 c^2}\int_0^\infty \dd\omega \, \omega^4 |\alpha_0(\omega)|^2\bkt{1-2\eta(\omega)}^2D^+(\omega)D^-(\omega)
    \label{eq:TD_deco_(2)}
\end{align}
As opposed to the Davies-Unruh decoherence coefficient, the time-dilation decoherence depends on the local acceleration both via the perceived field spectrum of the moving particle $  \bkt{\sim D^+ \bkt{\omega}D^- \bkt{\omega}}$, as well as the  direct scaling  $\sim a^2$ via the redshift factor. 

\subsection{Dispersion Potential}
\label{sec:CP_pot}
The terms $C^{(1)}_i$ and $C^{(2)}_{ij}$ represent derivatives of the dispersion potential arising from the interaction of the particle with the field. Using the split of the bath operator in Eq. (\ref{eq:bath_operator_split}) into a Davies-Unruh and a time-dilation contribution, we see that both $C^{(1)}_i$ and $C^{(2)}_{ij}$ will have contributions from both bath operators. The only nonvanishing contribution to \(C_i^{(1)}\) is due to the bath operator \(\hat B_i^\mr{TD}(\tau)\) of the time-dilation Hamiltonian \eqref{eq:H_int_TD}:
\eqn{
\label{eq:CP-force}
    C^{(1),\mr{TD}}_i=\frac{a_i}{2 \pi c^2}\int_0^\infty \dd\omega \, S(\omega)\omega^2\bkt{D^+(\omega)+D^-(\omega)}, 
}
where $S(\omega)$ is given by:
\begin{align}
    S(\omega)=\frac{1}{2}\re\left[\alpha_0(\omega)\eta(\omega)-\alpha_0 (\omega)\right].
\end{align}
$C^{(1),\mr{TD}}_i$ can be interpreted as the dispersion force arising from the effective temperature gradient across the extent of the particle's wavefunction. Similar corrections to the flat-space Casimir force between parallel plates that are linear in the local acceleration have been reported in \cite{Calloni_2002,Sorge_2005,Fulling_2007, Milton_2008, Sorge_2014, Sorge_2019}.

The second-order term \(C_i^{(2)}\) receives contributions from both \(\hat B_i^\mr{TD}(\tau)\) and the bath operator \(\hat B_i^\mr{DU}(\tau)\) defined by the Davies-Unruh Hamiltonian \eqref{eq:H_int_UD}. The Davies-Unruh contribution is
\begin{align}
C_{ij}^{(2),\mr{DU}} &= -\frac{\delta_{ij}}{24 \pi^2 \hbar c^2} \mc{P} \int_{-\infty}^\infty \dd \omega \int_{-\infty}^\infty \dd \omega' \frac{\omega^4 \omega'^2}{\omega + \omega'} |\alpha_0(\omega)|^2 \non \\
&\times \sbkt{D^+(\omega) D^+(\omega') - D^-(\omega) D^-(\omega')},
\end{align}
and the time-dilation contribution is
\begin{align}
&C_{ij}^{(2),\mr{TD}} = -\frac{a_i a_j}{32 \pi^2 \hbar c^4} \mc{P} \int_{-\infty}^\infty \dd \omega \int_{-\infty}^\infty \dd \omega' \frac{\omega^4}{\omega + \omega'} \non \\
&\times \vbkt{\alpha_0(\omega) [\eta(\omega) - 1]}^2 \sbkt{D^+(\omega) D^+(\omega') - D^-(\omega) D^-(\omega')}.
\end{align}
While $C_{ij}^{(2),\mr{DU}}$ represents an effect analogous to the thermal Lamb shift in atoms arising directly from the interaction of the particle with the effective thermal field, $C_{ij}^{(2),\mr{TD}}$ once again arises from the effective thermal gradient across the particles. The seeming contradiction of the appearance of a temperature gradient in an equilibrium force, such as $C_{i}^{(1),\mr{DU}}$ and $C_{ij}^{(2),\mr{DU}}$, can be related to the Tolman-Ehrenfest effect, which causes the field to equilibrate at a temperature gradient \cite{Tolman_1930, Ehrenfest_1930, Rovelli_2011}.

\section{Example: Hyperbolic motion}
\label{sec:hyperbolic_motion}
\subsection{Decoherence}
As a first example of decoherence in non-inertial reference frames, let us consider linear hyperbolic motion through Minkowski space. Without loss of generality we assume that our particle is accelerated along the x-axis with an acceleration $a$. The classical worldline of the particle can then be written as: 
\eqn{
t_0&=ca^{-1}\text{sinh}\bkt{a\tau/c}\non\\
x_0&=c^2a^{-1}\text{cosh}\bkt{a\tau/c}\non\\
y_0&=z_0=0
}
It was shown in \cite{Birrell_Davies_1982, Kim_Soh_Yee_1987} that the Fourier transforms of the forward and backward Wightman functions $D^+(\omega)$ and $D^-(\omega)$ pulled back onto a hyperbolic worldline are given by: 
\begin{align}
D^\pm(\omega) = -\frac{\hbar \omega}{2\pi \epsilon_0 c^3} \sbkt{\Theta(\mp \omega) + n(\omega)}, 
\end{align}
where $n(\omega)$ can be identified as the Bose-Einstein distribution at the Davies-Unruh temperature $T_\mr{DU}=\hbar a/(2\pi k_\mr{B}c)$, and $ \Theta \bkt{\omega} $ is the Heaviside function. To find all the decoherence coefficients, we substitute these Wightman functions into Eq. (\ref{eq:DU_deco})  and (\ref{eq:TD_deco_(2)}) to find: 
\begin{align}
    \Lambda^\mr{DU}=\frac{1}{3 (2\pi)^3\epsilon_0^2 c^8} \int_0^\infty \dd \omega \, \omega^8 |\alpha_0(\omega)|^2 \bkt{n(\omega) + 1} n(\omega),
\label{eq:deco_DU_linear_acc}
\end{align}
and

\begin{figure}[t]
    \centering
    \includegraphics[width=\linewidth]{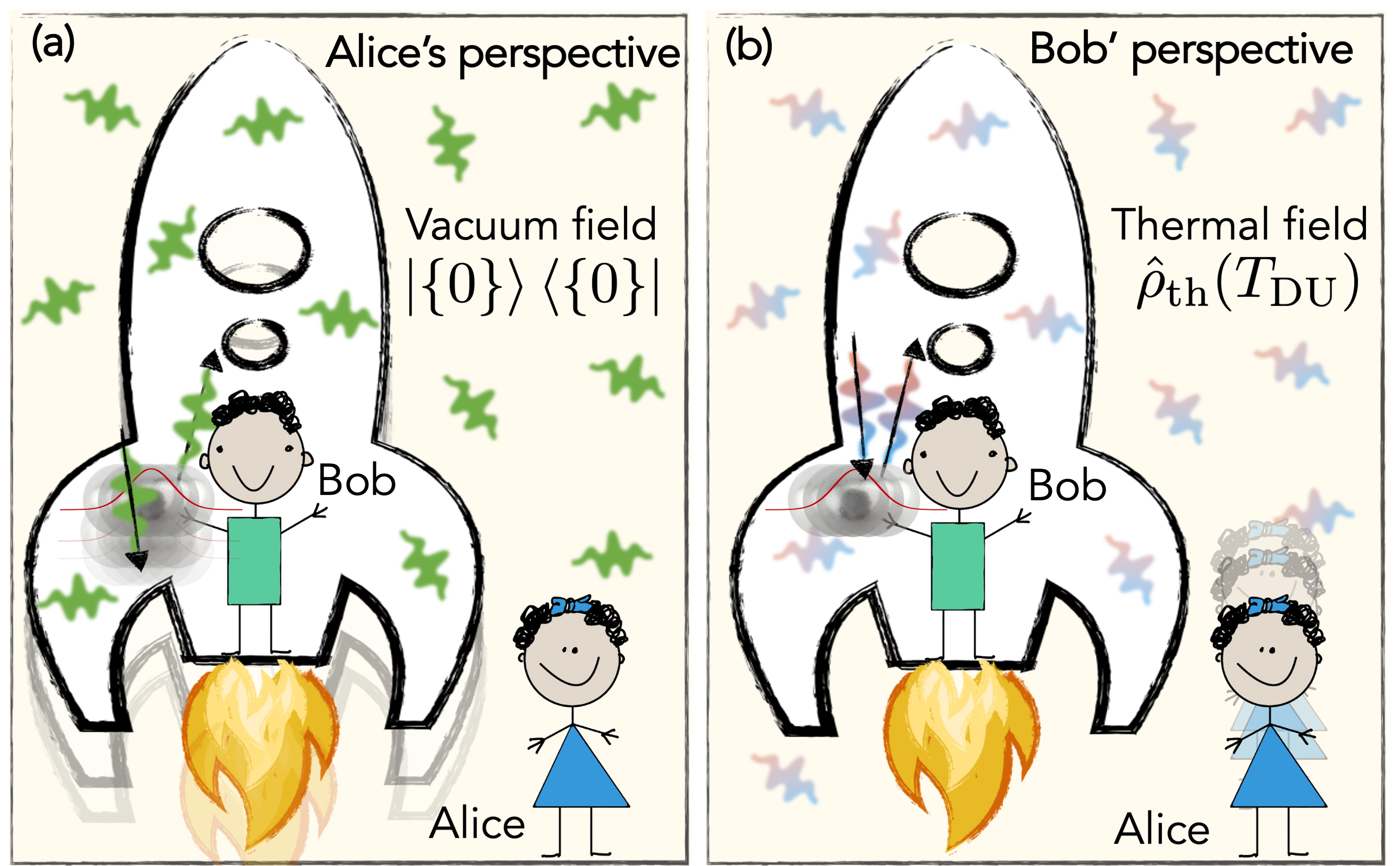}
    \caption{(a) An inertial observer, Alice, in Minkowski vacuum, observes the decoherence of a particle undergoing uniform acceleration, with Bob as a co-moving observer. From Alice's perspective the particle interacts with the zero-temperature vacuum field, scattering quantum fluctuations and experiencing momentum recoil in the process. This recoil is associated with `which-path' measurement, which leads to decoherence. (b) From Bob's perspective, the decoherence results from scattering by a thermal field.}
    \label{Fig:AliceBob}
\end{figure}

\begin{align}
    \Lambda_{ij}^\mr{TD}=&\frac{a_ia_j}{3(2\pi)^3\epsilon_0^2 c^4}\times\non\\
    &\int_0^\infty \dd\omega \, \omega^6 |\alpha_0(\omega)|^2\bkt{1-2\eta(\omega)}^2\bkt{n(\omega) + 1} n(\omega)
\end{align}

The Davies-Unruh coefficient Eq. \eqref{eq:deco_DU_linear_acc} takes on the same form as the decoherence coefficient Eq. \eqref{eq:thermal_deco} for a polarizable particle at rest interacting with a thermal field~\cite{PWM_KS22}. The numerical prefactor in the Davies-Unruh decoherence rate differs from that in \cite{Jakubec_2025}, due to the scalar nature of the field. We further note that the decoherence rate $ \Lambda^\mr{DU}$ is related to the momentum diffusion rate  of a particle undergoing hyperbolic motion  ~\cite{PWM_KS2024, PWM_KS22, Jakubec_2025}:
\eqn{ \frac{\avg{\Delta P^2 }}{\Delta t} = 2 \hbar^2 
\Lambda^\mr{DU}.
}
This can be understood as follows: the rate at which the effective thermal photons scatter off the particle and impart momentum kicks is the same rate at which they gain `which-path' information and decohere the particle's spatial superposition.
Both the decoherence coefficients $\Lambda^\mr{DU} $ and $\Lambda^\mr{TD} $ contain the factor $n(\omega)\bkt{n(\omega)+1} $, wherein the terms $ \sim n(\omega)$ and $ \sim n^2(\omega)$ correspond to the particle-like and wave-like contributions of the thermal field, respectively~\cite{PWM_KS22}. The appearance of the Bose-Einstein distribution also ensures the convergence of the integral.

Additionally,  the decoherence of the particle undergoing hyperbolic motion can be understood from the perspective of an inertial observer in the Minkowski vacuum as follows. Consider Alice to be an inertial observer, while Bob is located in the co-moving frame with the particle (see Fig.~\ref{Fig:AliceBob}).  Since the field in Bob's frame appears thermal, the decoherence in Bob's frame arises from the scattering of apparent thermal photons by the particle~\cite{Joos1985, schlosshauer2007decoherence}. In Alice's frame,  as the quantum fluctuations of the  vacuum field interact with the accelerating particle and scatter, the particle experiences recoil. This recoil results in both a momentum diffusion, and concomitantly decoherence in the position basis. We note that such an effect is consistent with the fact that Alice does not measure any excitations in the field~\cite{RSG, FordOConnell}.

\subsection{Dispersion Force}
Using the same $D^+(\omega)$ and $D^-(\omega)$, we can also calculate the dispersion force experienced by the particle. Substituting the Wightman functions into Eq. (\ref{eq:CP-force}), we get: 
\begin{align}
     C^{(1),\mr{TD}}_i=&\frac{\hbar a_i}{8\pi^2 \epsilon_0c^5}\times\non\\
     &\int_0^\infty \dd\omega \, \omega^3 \re\sbkt{\alpha_0(\omega)(\eta(\omega) - 1)}\bkt{2n(\omega)+1)}, 
\end{align}
where $n(\omega)$ is once again the Bose-Einstein distribution at the Davies-Unruh temperature. As explained in Section \ref{sec:CP_pot}, this force is related to the temperature gradient experienced by the particle across its wavefunction. 
\section{Example: Circular Motion}
\label{sec:circular_motion}
\subsection{Decoherence}
The second example we consider is that of relativistic uniform circular motion. Circular motion with relativistic speed is commonly achieved in accelerators and has been studied before in \cite{Bell_1983, Bell_1987, Goyal_2023}. The worldline of a particle undergoing circular motion parameterized by its proper time $\tau$ is given by:  
\begin{align}
    t_0&=\tau\gamma\non\\
    x_0&=\rho\mr{sin}\bkt{\xi\tau\gamma}\non\\
    y_0&=\rho\mr{cos}\bkt{\xi\tau\gamma}\non\\
    z_0&=0
\end{align}
Here $\rho$ is the radius, $\gamma=1/\bkt{1-v^2/c^2}^\frac{1}{2}$, $\xi$ is the angular frequency and $v=\rho\xi$ is the speed of the particle. The local acceleration experience by the particle in its rest frame is: 
\begin{align}
    a=v\xi\gamma.
\end{align}
To find the decoherence coefficient, we will focus on the ultra-relativistic case $v\approx c$. In this case, the Fourier transform of the Wightman function can be simplified and has been found in \cite{Kim_Soh_Yee_1987} to be:
\begin{align}
    D^+(\omega) &\simeq \frac{\hbar \xi}{8\pi \epsilon_0 c^3 R \gamma} \sbkt{\frac{e^{-2\omega R/\xi \gamma}}{(v/c) \cosh(R) - 1}} \\
    D^-(\omega) &\simeq \frac{\hbar \omega}{2\pi \epsilon_0 c^3} + \frac{\hbar \xi}{8\pi \epsilon_0 c^3 R \gamma} \sbkt{\frac{e^{-2\omega R/\xi \gamma}}{(v/c) \cosh(R) - 1}}
\end{align}
where $R$ is the positive real root of the equation \(R = (v/c) \sinh R\). For simplicity we define a function:
\begin{align}
    C(R)=\frac{\xi}{4\pi\gamma}\frac{1}{2R\bkt{(v/c) \ \mr{cosh}R-1}}.
\end{align}
Substituting these expressions into Eqs. (\ref{eq:DU_deco})  and \eqref{eq:TD_deco_(2)} gives us: 
\begin{widetext}
    \begin{align}
        \Lambda^\mr{DU}&=\frac{1}{6\pi \epsilon_0^2c^4}\int_0^\infty \dd\omega \, \omega^6 |\alpha_0(\omega)|^2 C(R)\sbkt{C(R)e^{-2\omega R/\xi\gamma}+\frac{\omega}{2\pi}}e^{-2\omega R/\xi\gamma},\\
        \Lambda_{ij}^\mr{TD}&=\frac{a_ia_j}{6\pi\epsilon_0^2 c^4}\int_0^\infty \dd\omega \, \omega^4 |\alpha_0(\omega)|^2\bkt{1-2\eta(\omega)}^2C(R)\sbkt{C(R)e^{-2\omega R/\xi\gamma}+\frac{\omega}{2\pi}}e^{-2\omega R/\xi\gamma}.
    \end{align}
\end{widetext}
The decaying exponential in the integrand once again ensures the convergence of the frequency integral. Compared to hyperbolic motion, the Bose-Einstein distribution is replaced by a distribution determined by the factor $C(R)$.
\subsection{Dispersion Force}
We again find the dispersion force by substituting the Wightman functions for circular motion into Eq. (\ref{eq:CP-force}), which yields:
\begin{align}
    C_i^{(1),\mr{TD}}=&\frac{\hbar a_i}{8 \pi^2 \epsilon_0c^5}\int_0^\infty \dd\omega \, \omega^2\re\sbkt{\alpha_0(\omega)(1-\eta(\omega))}\times\non\\
    &\bkt{\frac{\omega}{4\pi}+C(R)e^{-2\omega R/\xi\gamma}}.
\end{align}
For the case where the particle moves along a worldline with a constant radius, the local acceleration is the centrifugal force  pointing radially outwards. The dispersion force thus becomes: 
\begin{align}
    C_i^{(1),\mr{TD}}=&\frac{\hbar \xi^2\gamma^2\rho}{8 \pi^2 \epsilon_0c^5}\int_0^\infty \dd\omega \, \omega^2\re\sbkt{\alpha_0(\omega)(1-\eta(\omega))}\times\non\\
    &\bkt{\frac{\omega}{4\pi}+C(R)e^{-2\omega R/\xi\gamma}}.
\end{align}
Like the decoherence coefficient, the field spectrum is described by a distribution determined by $ C(R)$. 
\section{Summary and Outlook}
\label{sec:discussion_outlook}
In this paper we analyze the open quantum system dynamics of the center of mass of a point-like particle coupled to a scalar field, and moving along a stationary worldline through 3+1 dimensional Minkowski space. In our model the particle has two distinct degrees of freedom: (1) an internal oscillator  that couples to the field, and   (2) its quantized center-of-mass motion, which is considered to be spatially delocalized around its worldline. We first  derive an effective red-shifted polarizability of the particle undergoing non-inertial motion by solving the  coupled dynamics of the  internal oscillator and the field (Eq.~\eqref{eq:redshift_polarizability}). 
We then describe the open system dynamics of the particle's quantized center-of-mass motion via a QBM master equation, derived under the Born-Markov approximation (Eq.~\eqref{eq:QBM_master_equation}). The resulting QBM master equation   consists of (1) a dispersive contribution corresponding to a center-of-mass Hamiltonian modification (Eqs.~\eqref{eq:c1} and \eqref{eq:c2}), akin to Casimir-Polder forces on particles near media~\cite{Jakubec_2025}; and (2) dissipative terms that include center-of-mass decoherence in the position basis. 
For stationary worldlines, we find that the dispersive potential seen by the particle scales linearly with the local acceleration $a_i$, resembling the gravitational corrections to Casimir forces found in \cite{Calloni_2002,Sorge_2005,Fulling_2007, Milton_2008, Sorge_2014, Sorge_2019}. 

We highlight two distinct sources of decoherence arising from two distinct system-bath coupling mechanism: 

\begin{enumerate}
    \item \textit{Davies-Unruh decoherence}: A particle moving along a stationary worldline registers a modified field spectrum, which differs from that of the Minkowski vacuum seen by an inertial observer. When this field interacts with the particle and imparts recoil, which-path information about the particle's center-of-mass position is acquired, and this manifests itself as decoherence.  Such decoherence has  a form that is effectively thermal, Eq.~\eqref{eq:DU_deco}. This is further confirmed by the fact that in the special case of hyperbolic motion, the decoherence rate of a uniformly accelerating particle (Eq.~\eqref{eq:deco_DU_linear_acc})  is equivalent to that of a particle at rest immersed in a thermal field at temperature $ T_\mr{DU} = \hbar a /(2\pi k_\mr{B} c)$~\cite{Jakubec_2025, PWM_KS22}. Additionally, we find that the same decoherence coefficient can be arrived at by considering the thermal decoherence rate of a polarizable point-particle in flat spacetime (Eq.~\eqref{eq:thermal_deco}), and imposing the condition that the effective field observed by the particle on a stationary worldline satisfies the detailed balance condition (Eq.~\eqref{eq:detailed_balance_stationary}).  
    \item \textit{Time-dilation decoherence}: Since the particle contains an internal oscillator with a characteristic frequency, it is sensitive to time-dilation across its delocalized center of mass. A spatial superposition parallel to the acceleration experienced by the particle leads to a differential time dilation across the  particle's wavefunction. This induces an effective coupling between the center of mass and the field, resulting in decoherence. Along a stationary worldline, such time-dilation decoherence (Eq. \eqref{eq:TD_deco_(2)}) is also thermal in nature, represented by the appearance of the factor $D^+(\omega)D^-(\omega)$, but with an explicit dependence on the local acceleration as well as on the particle's red-shifted polarizability (Eq.~\eqref{eq:redshift_polarizability}). 
\end{enumerate}

 We  evaluate both the dispersion force and  decoherence rates for two specific examples of stationary worldlines: hyperbolic motion (Section \ref{sec:hyperbolic_motion}) and circular motion (Section \ref{sec:circular_motion}). One of the distinguishing features between these two types of motion is that hyperbolic motion gives rise to a Rindler horizon, whereas circular motion does not. 
  Considering the recent interest in  decoherence near black holes \cite{DSW20222, DSW_Killing, Wei2024, Chen2024, Ordonez01012026} and the role played by the presence of horizons in engendering it \cite{Batista_2026}, we remark that in our model we find a finite decoherence rate for both hyperbolic and circular motion. Furthermore, we note that there are striking similarities between the Davies-Unruh decoherence rate of a particle undergoing non-inertial motion  (Eq.~\eqref{eq:DU_deco}), and the decoherence  of a polarizable particle at rest near a medium, immersed in a thermal field~\cite{Sinha2020PRA, biggs2024, Jakubec_2025}.  This correspondence motivates further investigations into the open system dynamics of test quantum systems undergoing non-inertial motion, or equivalently in a curved spacetime, suggesting a characterization of curved spacetime as an effective reservoir. 

\section{Acknowledgments}
We are grateful to Sam Gralla and Hongji Wei for helpful discussions. C.J. would like to thank Uro\v{s} Deli\'{c} for his support. This work was supported by the National Science Foundation under Grant No. PHY-2418249,   the John Templeton Foundation under Award No. 63626, and the U.S. Department of Energy, Office of Science under Grant No. DESC0026059. C. J. acknowledges support by the John Templeton Foundation under Award No. 63033.

\begin{widetext}
\appendix
\section{System Lagrangian and Hamiltonian}
\label{App:Lag+Ham}
We start with a particle whose center-of-mass moves along a prescribed classical worldline $x_0^\mu(\tau)$ parameterized by the proper time $\tau$. We describe the \textit{spatial} quantized center-of-mass fluctuations of the particle around $x_0^\mu(\tau)$ in the local rest frame of the particle and denote them by $\hat{X}^i$. The conjugate momentum in the local rest frame is $\hat P_i$ and we impose the canonical commutation relation $\sbkt{\hat X^i,\hat P_j}=i\hbar\delta^i_{j}$.
Following the description in \cite{Pikovski2015, Zych_Thesis}, we can write the Lagrangian in the local rest frame as:
\begin{align}
L=L_\mr{rest}\sbkt{g_{00}(\hat X^i)-g_{mn}(\hat X^i)\dot{\hat X}^m\dot{\hat X}^n/c^2}^{\frac{1}{2}},
\end{align}
where for simplicity we have assumed that $g_{0m}=g_{m0}=0$. The rest Lagrangian $L_\mr{rest}$ is given by: 
\begin{align}
L_\mr{rest}=-\mathbb{1} \ Mc^2+\frac{m\dot {\hat {q}}(\tau)^2}{2}-\frac{m\omega_q^2\hat q(\tau)^2}{2}.
\end{align}
Here $M$ is the  mass of the particle, $\hat q(\tau)$ is the internal degree of freedom position, which is assumed to be point-like and thus does not transform under general coordinate transformations, $\omega_q$ is the frequency in the rest frame of the particle, and $m$ is the mass associated with the internal oscillator. Using the Lagrangian in the local rest frame, we can calculate the Hamiltonian in the local rest frame by taking the Legendre transform with respect to both the internal and  center-of-mass degrees of freedom, which yields: 
\begin{align}
\hat H=\sqrt{g_{00}(\hat X^i)\bkt{\hat E_\mr{rest}^2+\hat P^i\hat P_ic^2}}
\end{align}
We note that the metric $g_{\mu\nu}$ which appears in the form of the red-shift factor $g_{00}$ and in the inner product of $\hat{P}^i$ as $g_{ij}$ is evaluated around the particle worldline $x_0^\mu(\tau)$ and can thus be expanded in the local rest frame in terms of $\hat X^i$. Finally, we assume that the particle moves non-relativistically in the  local rest frame such that the kinetic contribution to total system energy is much smaller than the rest energy contribution. In other words, we assume that there are no special relativistic time-dilation effects in the local rest frame of the particle. We then end up with Eq. (\ref{eq:Hs_NR}).

\section{Internal oscillator dynamics and  polarizability}

\subsection{Coupled dynamics of the internal oscillator and the field}
\label{App:intfield}

In order to describe the response of the particle to the external field, we will now derive an effective polarizability of the particle, defined through its linear response to the field.
%
Using Eq.~\eqref{eq:Hs_NR}, the Heisenberg equations of motion for the internal oscillator are given by:
\begin{align}
\label{eq:eom_x}
    \dot{\hat q}&=-\frac{\ii}{\hbar}\sbkt{\hat q, \hat H_\mr S}=\frac{\sqrt{-g_{00}}}{m}\bkt{ \hat p-e\hat \phi(\hat X^i,\tau)}\\
    \dot{\hat p}&=-\frac{\ii}{\hbar}\sbkt{\hat p, \hat H_\mr S}=-\sqrt{-g_{00}} \, m\omega_q^2\hat q.
\end{align}
Combining these two equations leaves us with: 
\begin{align}
\label{eq:B3}
    m\ddot{\hat q}-g_{00}m\omega_q^2\hat q^2=-\sqrt{-g_{00}}{e}\dot{\hat \phi}(\hat X^i,\tau),
\end{align} 
from which we can see that the time derivative of the field acts as a driving force on the internal oscillator. 

Similarly, the Heisenberg equation of motion for the field mode operators are: 
\begin{align}
    \dot{\hat g}_\mb{k}=-\frac{\ii}{\hbar}\sbkt{\hat g_\mb{k}, \hat H_\mr F+\hat H_\mr S}=- \ii\Omega_k\hat g_\mb{k}+\frac{\ii e \sqrt{- g_{00}}}{\hbar m}\bkt{\hat p-e\hat \phi(\hat X^i,\tau)}\phi^*_\mb{k}(\hat X^i,\tau),
\end{align}
which we solve to get:
\begin{align}
    \hat g_\mb{k}\bkt{ \tau}=\hat g_\mb{k}\bkt{0}e^{-\ii\Omega_\mb{k} \tau} + \frac{\ii e}{\hbar}\int_0^\tau \dd \tau' \der{\hat q \bkt{\tau'}}{\tau'} \phi_\mb{k}^*(\hat X^i, \tau') e^{-\ii\Omega_\mb{k} (\tau-\tau')}.
\end{align}
Therefore the field is given by:
\begin{align}
\label{eq:B7}
\hat\phi_\mr{tot}(\hat X^i,\tau) &= \hat \phi(\hat X^i,\tau)\non\\
&+ \frac{\ii e}{\hbar}\int_0^\tau \dd\tau' \der{\hat q \bkt{\tau'}}{\tau'}\sum_\mb{k}\sbkt{\phi_\mb{k}(\hat X^i, \tau) \phi^*_\mb{k}(\hat X^i, \tau')e^{-\ii\Omega_\mb{k}(\tau-\tau')}-\phi_\mb{k}^*(\hat X^i,\tau)\phi_\mb{k}(\hat X^i, \tau')e^{\ii\Omega_\mb{k}(\tau-\tau')}}\\
&=\hat \phi(\hat X^i,\tau)\underbrace{+\frac{\ii e}{\hbar}\int_0^\tau \dd\tau ' \der{\hat q\bkt{\tau'}}{\tau'} \sbkt{\avg{\hat \phi(\hat X^i,\tau)\hat \phi(\hat X^i,\tau')}-\avg{\hat \phi(\hat X^i,\tau')\hat \phi(\hat X^i,\tau)}}}_{\hat\phi_\mr{rr}(\hat X^i,\tau)}
\end{align}
where $\hat\phi(\hat X^i,\tau)$ is the free-evolving field (i.e., the part evolved by $\hat H_\mr{F}$). The radiation-reaction part $\hat\phi_\mr{rr}(\hat X^i,\tau)$ can be simplified by taking the Fourier transform with respect to $\tau'$, using the definitions of $D^\pm\bkt{\Omega}$ in Eq.~\eqref{eq:FT_correlators} and neglecting $\mathcal{O}((\hat X^i)^2)$ terms:
\eqn{
\hat\phi_\mr{rr}(\tau) &\simeq \frac{\ii e}{2\pi \hbar} \int_0^\tau \dd \tau' \int_{-\infty}^\infty \dd \Omega \der{\hat q(\tau')}{\tau'} \sbkt{D^+(\Omega) - D^-(\Omega)} e^{\ii \Omega(\tau-\tau')}\\
&= \frac{e}{2\pi^2 \epsilon_0 c^3} \int_0^\tau \dd \tau' \int_0^\infty \dd \Omega \der{\hat q(\tau')}{\tau'} \Omega \sin(\Omega(\tau-\tau')),
}
The time derivative of the radiation-reaction term is therefore: 
\begin{align}
\label{eq:B13}
\dot{\hat\phi}_\mr{rr}(\tau) &= \frac{e}{2\pi^2 \epsilon_0 c^3} \int_0^\tau \dd \tau' \int_0^\infty \dd \Omega \der{\hat q\bkt{\tau'}}{\tau'} \Omega^2 \cos(\Omega(\tau-\tau'))  \\
&= -\frac{e}{2\pi^2 \epsilon_0 c^3} \int_0^\tau \dd \tau' \int_0^\infty \dd \Omega \der{\hat q(\tau')}{\tau'} \partial_\tau^2 \cos(\Omega(\tau-\tau'))  \\
&= -\frac{e}{2\pi \epsilon_0 c^3} \int_0^\tau \dd \tau' \der{\hat q (\tau')}{\tau'} \partial_\tau^2 \delta(\tau-\tau')  \\
&= -\frac{e}{4\pi \epsilon_0 c^3}\bkt{ \der{^3 \hat q \bkt{\tau}}{\tau^3}-\der{^2 \hat q \bkt{\tau}}{\tau^2}\delta(0)}.
\end{align}
Then from Eqs. (\ref{eq:B3}), (\ref{eq:B7}), and (\ref{eq:B13}) we get: 
\begin{align}\label{eq:qdd}
(m+\delta m)\ddot{\hat q} - g_{00}m\omega_q^2 \hat q - \sqrt{-g_{00}} \frac{e^2}{4\pi \epsilon_0 c^3} \dddot{\hat q} = - \sqrt{-g_{00}} e \dot{\hat \phi}(\hat X^i,\tau),.
\end{align}
$\delta m$ is a contribution to the particle mass resulting from the coupling of the particle to its own radiation reaction field, and is given by the divergent expression $(e^2/4\pi\epsilon_0c^3)\delta(0)$ in our calculation assuming a point-like particle. To be consistent with our nonrelativistic approximation for the internal dynamics of the particle we can replace $\delta(0)$ by $\Omega/\pi$, where $\Omega$ is a high-frequency cutoff; this results in the replacement of $\delta m/m$ by 
\begin{align}
\frac{\delta m}{m}=\frac{e^2\Omega}{4\pi^2\epsilon_0mc^3}=\frac{r_0}{\pi c}\Omega,
\end{align}
where $r_0=e^2/(4\pi\epsilon_0mc^2)\cong 2.8\times 10^{-15} m$ is the classical electron radius. Thus $\delta m/m$ may be assumed to be negligible for any cutoff frequency $\Omega$ much smaller than $10^{23}$ Hz, and then we can ignore the mass renormalization in Eq. (\ref{eq:qdd}).

\subsection{Effective polarizability}
\label{App:polarizability}
Taking the Fourier transform of Eq.~\eqref{eq:qdd} allows us to write the steady state induced dipole $\hat d_\mr{st}$ as: 
\begin{align}
    \hat{d}_\mr{st}(\hat X^i,\tau)=e\hat q(\hat X^i,\tau)=-\frac{\ii}{2\pi}\sqrt{-g_{00}}\int \dd\omega \, \omega\alpha(\omega)\underaccent{\bar}{\hat\phi}(\hat X^i,\omega) e^{\ii \omega \tau}, 
\end{align}
where $\alpha(\omega)$ is the polarizability of the particle, which describes its response to an external field: 
\begin{align}
    \alpha(\omega) \coloneqq \frac{e^2/ m}{2\ii \sqrt{-g_{00}} \beta \omega^3 - \omega^2 - g_{00} \omega_q^2}; \qquad \beta \coloneqq \frac{e^2/m}{8\pi \epsilon_0 c^3}.
\end{align}
 From this expression we can see that the characteristic frequencies that determine the response of the particle are red-shifted across the spread of the wavefunction, which leads to a coupling between the the quantized center of mass and the field. This can be seen by expanding the polarizability to first order in $\hat X^i$ using Eq.~\eqref{eq:FW_metric_00} to get: 
\begin{align}
    \alpha(\omega)\approx \alpha_0(\omega) \sbkt{1 - \eta(\omega) a_i \hat X^i/c^2} .
\end{align}
Here $\alpha_0(\omega)$ is the bare polarizability in the rest frame of the particle and $\eta(\omega)$ is a redshift-induced correction (see Eqs.~\eqref{eq:alpha_0} and \eqref{eq:eta} respectively). Combining Eq.~\eqref{eq:Erest}, Eq.~\eqref{eq:eom_x} and Eq.~\eqref{eq:redshift_polarizability} we can rewrite Eq.~\eqref{eq:Hs_NR} in the Schr\"odinger picture as given by Eq.~\eqref{eq:hs}. 

\section{Decoherence rate}

\label{App:decoherence_coefficients}
\subsection{Davies-Unruh Decoherence from First Principles}
The derivation of $\Lambda^\mr{DU}_{ij}$, with $i=j$ corresponding to the Davies-Unruh decoherence coefficient, goes as follows. Using the bath operator $B_i^\mr{DU}(\tau)$ in the interaction picture (Eq.~\eqref{eq:Bdu}): 
\begin{align}
\Lambda^\mr{DU}_{ij}=&\frac{1}{2\hbar^2}\int_0^\infty \dd\tau'\left\langle\cbkt{\hat B_i^\mr{DU}(\tau),\hat B_j^\mr{DU}(\tau-\tau')}\right\rangle\\
=&\frac{1}{2\hbar^2}\int_0^\infty \dd\tau' \left\langle \hat B_i^\mr{DU}\bkt{\tau}\hat B_j^\mr{DU}(\tau-\tau')\right\rangle + \frac{1}{2\hbar^2}\int_0^\infty \dd\tau' \left\langle \hat B_j^\mr{DU}(\tau-\tau')\hat B_j^\mr{DU}(\tau)\right\rangle.
\end{align}
Substituting the expressions for the Davies-Unruh bath operator gives: 
\begin{align}
\Lambda^\mr{DU}_{ij}=\frac{1}{2\hbar^2}\int_0^\infty \dd\tau' \left\langle \dot{\hat d}_{0,\mr{st}}(\tau)\partial_i\hat \phi(\tau) \dot{\hat d}_{0,\mr{st}}(\tau-\tau')\partial_j\hat \phi(\tau-\tau') \right\rangle + \frac{1}{2\hbar^2}\int_0^\infty \dd\tau' \left\langle \dot{\hat d}_{0,\mr{st}}(\tau-\tau')\partial_i\hat \phi(\tau-\tau') \dot{\hat d}_{0,\mr{st}}(\tau)\partial_j\hat \phi(\tau) \right\rangle,
\end{align}
where we use the shorthand notation $\hat \phi(\tau)\equiv \hat \phi(x_0^\mu(\tau))$. In the next step we apply Wick's theorem to break up the 4-point correlators into products of 2-point correlators. Additionally, we make use of our assumption that the field and its first derivative are statistically independent. Together these assumptions yield: 
\begin{align}
\Lambda^\mr{DU}_{ij}=&\frac{1}{2\hbar^2}\int_0^\infty \dd\tau'\left\langle \dot{\hat d}_{0,\mr{st}}(\tau)\dot{\hat d}_{0,\mr{st}}(\tau-\tau')\right\rangle\left\langle \partial_i\hat \phi(\tau)\partial_j\hat \phi(\tau-\tau')\right\rangle\non\\
&+
\frac{1}{2\hbar^2}\int_0^\infty \dd\tau'\left\langle \dot{\hat d}_{0,\mr{st}}(\tau-\tau')\dot{\hat d}_{0,\mr{st}}(\tau)\right\rangle\left\langle \partial_i\hat \phi(\tau-\tau')\partial_j\hat \phi(\tau)\right\rangle.
\end{align}
Now we use the assumption that different spatial directions of the center-of-mass are uncorrelated. This implies that $\langle\partial_j\hat\phi(\tau)\partial_i\hat\phi (\tau-\tau')\rangle=0$ and leaves us with: 
\begin{align}
\Lambda^\mr{DU}_{ii}=&\frac{1}{2\hbar^2}\int_0^\infty \dd\tau'\left\langle \dot{\hat d}_{0,\mr{st}}(\tau)\dot{\hat d}_{0,\mr{st}}(\tau-\tau')\right\rangle\left\langle \partial_i\hat \phi(\tau)\partial_i\hat \phi(\tau-\tau')\right\rangle\non\\
&+
\frac{1}{2\hbar^2}\int_0^\infty \dd\tau'\left\langle \dot{\hat d}_{0,\mr{st}}(\tau-\tau')\dot{\hat d}_{0,\mr{st}}(\tau)\right\rangle\left\langle \partial_i\hat \phi(\tau-\tau')\partial_i\hat \phi(\tau)\right\rangle.
\end{align}
Since $T_\mr{eff}$ is a function of $\sqrt{a_ia^i}$ and thus isotropic, we assume that $\Lambda^\mr{DU}_{ij}$ is independent of which axis the superposition is prepared in. In combination with $\langle\partial_j\hat\phi(\tau)\partial_i\hat\phi (\tau-\tau')\rangle=0$ this allows us to simplify the correlator of the spatial derivatives: 
\begin{align}
\label{eq:grad_phi_grad_phi}
\left\langle \partial_i\hat \phi(X)\partial_i\hat \phi(X')\right\rangle=\frac{1}{3c^2}\left\langle \partial_\tau\hat \phi(X)\partial_\tau\hat \phi(X')\right\rangle.
\end{align}
After this, $\Lambda^\mr{DU}$ is: 
\begin{align}
\label{eq:C7}
\Lambda^\mr{DU}=&\frac{1}{6\hbar^2c^2}\int_0^\infty \dd\tau'\left\langle \dot{\hat d}_{0,\mr{st}}(\tau)\dot{\hat d}_{0,\mr{st}}(\tau-\tau')\right\rangle\left\langle \partial_\tau\hat \phi(\tau)\partial_\tau\hat \phi(\tau-\tau')\right\rangle\non\\
&+
\frac{1}{6\hbar^2c^2}\int_0^\infty \dd\tau'\left\langle \dot{\hat d}_{0,\mr{st}}(\tau-\tau')\dot{\hat d}_{0,\mr{st}}(\tau)\right\rangle\left\langle \partial_\tau\hat \phi(\tau-\tau')\partial_\tau\hat \phi(\tau)\right\rangle.
\end{align}
We now write the field and induced polarization in the interaction picture in terms of their spectral decomposition: 
\begin{align}
\hat \phi(\tau)&=\frac{1}{2\pi}\int_{-\infty}^{\infty}\dd\omega \, \underaccent{\bar}{\hat\phi}(\omega,\tau),\non\\
\hat d_{0,\mr{st}}(\tau)&=\frac{\ii}{2\pi}\int_{-\infty}^\infty \dd\omega \, \omega\alpha_0(\omega)\underaccent{\bar}{\hat \phi}( \omega,\tau),
\end{align}
which yields:
\begin{align}
\Lambda^\mr{DU}=&\frac{1}{(2\pi)^26\hbar^2c^2}\int_0^\infty \dd\tau'\int_{-\infty}^{\infty} \dd\omega_1 \, \omega_1^4 |\alpha_0(\omega_1)|^2\left\langle \underaccent{\bar}{\hat \phi}(\omega_1,\tau)\underaccent{\bar}{\hat \phi}(\omega_1,\tau-\tau')\right\rangle\int_{-\infty}^{\infty}\dd\omega_2 \, \omega_2^2\left\langle \underaccent{\bar}{\hat \phi}(\omega_2,\tau)\underaccent{\bar}{\hat \phi}(\omega_2,\tau-\tau')\right\rangle\non\\
+&\frac{1}{(2\pi)^26\hbar^2c^2}\int_0^\infty \dd\tau'\int_{-\infty}^{\infty} \dd\omega_1 \, \omega_1^4 |\alpha_0(\omega_1)|^2\left\langle \underaccent{\bar}{\hat \phi}(\omega_1,\tau-\tau')\underaccent{\bar}{\hat \phi}(\omega_1,\tau)\right\rangle\int_{-\infty}^{\infty}\dd\omega_2 \, \omega_2^2\left\langle\underaccent{\bar}{ \hat \phi}(\omega_2,\tau-\tau')\underaccent{\bar}{\hat \phi}(\omega_2,\tau)\right\rangle.
\end{align}
In the second term we now make the transformation $\tau' \rightarrow -\tau'$, which will cause the correlators to change:
\begin{align}
\left\langle \underaccent{\bar}{\hat \phi}(\omega,\tau-\tau')\underaccent{\bar}{\hat \phi}(\omega,\tau)\right\rangle \rightarrow \left\langle \underaccent{\bar}{\hat \phi}(\omega,\tau+\tau')\underaccent{\bar}{\hat \phi}(\omega,\tau)\right\rangle=\left\langle \underaccent{\bar}{\hat \phi}(\omega,\tau)\underaccent{\bar}{\hat \phi}(\omega,\tau-\tau')\right\rangle, 
\end{align}
where in the last equality we have used the fact that the pullback of the correlator onto the worldline of the particle only depends on $\tau'$; this follows from the stationarity of the particle's worldline. Thus the correlation function is invariant under proper time translations along the worldline. With this we can simplify the decoherence expression to: 
\begin{align}
\Lambda^\mr{DU}=\frac{1}{(2\pi)^26\hbar^2c^2}\int_{-\infty}^\infty \dd\tau'\int_{-\infty}^{\infty}\dd\omega_1 \, \omega_1^4 |\alpha_0(\omega_1)|^2\left\langle \underaccent{\bar}{\hat \phi}(\omega_1,\tau)\underaccent{\bar}{\hat \phi}(\omega_1,\tau-\tau')\right\rangle\int_{-\infty}^{\infty}\dd\omega_2 \, \omega_2^2\left\langle \underaccent{\bar}{\hat \phi}(\omega_2,\tau)\underaccent{\bar}{\hat \phi}(\omega_2,\tau-\tau')\right\rangle.
\end{align}
Remembering that the field operators are in the interaction picture,  we can write: 
\eqn{
\Lambda^\mr{DU}&=\frac{1}{(2\pi)^26\hbar^2c^2}\int_{-\infty}^\infty \dd\tau'\int_{-\infty}^{\infty}\dd\omega_1\int_{-\infty}^{\infty}\dd\omega_2 \, \omega_1^4\omega_2^2 |\alpha_0(\omega_1)|^2 D^+(\omega_1)D^+(\omega_2)e^{-i(\omega_1+\omega_2)\tau'}.
}
Simplifying further:
\eqn{
\Lambda^\mr{DU}&=\frac{1}{12\pi\hbar^2c^2}\int_{-\infty}^{\infty}\dd\omega \, \omega^6 |\alpha_0(\omega)|^2 D^+(\omega)D^-(\omega)\\
&=\frac{1}{6\pi\hbar^2c^2}\int_{0}^{\infty}\dd\omega \, \omega^6 |\alpha_0(\omega)|^2 D^+(\omega)D^-(\omega),
}
which is our final form of the Davies-Unruh decoherence coefficient. Here $D^+(\omega)$ and $D^-(\omega)$ are the Fourier transforms of the pullback of the forward and backward processes respectively. 

\subsection{Time Dilation Decoherence}

As before, we start by writing the time-dilation decoherence coefficient as: 
\begin{align}
\label{eq:Lambda_TD_bath_op}
\Lambda^\mr{TD}_{ij}=&\frac{1}{2\hbar^2}\int_0^\infty \dd\tau'\left\langle\cbkt{\hat B^\mr{TD}_i(\tau),\hat B^\mr{TD}_j(\tau-\tau')}\right\rangle\non\\
=&\frac{1}{2\hbar^2}\int_0^\infty \dd\tau' \left\langle \hat B^\mr{TD}_i(\tau)\hat B^\mr{TD}_j(\tau-\tau')\right\rangle + \frac{1}{2\hbar^2}\int_0^\infty \dd\tau' \left\langle \hat B^\mr{TD}_j(\tau-\tau')\hat B^\mr{TD}_j(\tau)\right\rangle
\end{align}
Comparing the bath operators $\hat{B}_i^\mr{DU}(\tau)$ and $\hat{B}_i^\mr{TD}(\tau)$ (Eqs.~\eqref{eq:Bdu} and \eqref{eq:Btd}), we can obtain immediately the time-dilation decoherence coefficient by realizing that it is of the same structure as Eq. (\ref{eq:C7}), except with red-shifted polarizabilities: 
\begin{align}
    \Lambda_{ij}^\mr{TD}
    =\frac{a_ia_j}{6\pi\hbar^2 c^2}\int_0^\infty \dd\omega \, \omega^4 |\alpha_0(\omega)|^2 \bkt{1-2\eta(\omega)}^2D^+(\omega)D^-(\omega).
\end{align}

\section{Second-order dispersion potential}
Here we compute the second-order contributions to the dispersion potential by substituting the bath operators \(\hat B_i^\mr{DU}(\tau)\) and \(\hat B_i^\mr{TD}(\tau)\) (Eqs. \eqref{eq:Btd} and \eqref{eq:Bdu}) into Eq. \eqref{eq:c2}.

\label{App:dispersion_potential}
\subsection{Davies-Unruh Dispersion Potential}
The Davies-Unruh term of the dispersion potential is
\eqn{
C_{ij}^{(2),\mr{DU}} &= \frac{\ii}{2\hbar} \int_0^\infty \dd \tau' \avg{\sbkt{\hat B_i^\mr{DU}(\tau), \hat B_j^\mr{DU}(\tau-\tau')}}.
}
As done in the decoherence calculations, we substitute \(\hat B_i^\mr{DU}(\tau)\) and split the 4-point correlators into 2-point correlators,
\begin{align}
C_{ij}^{(2),\mr{DU}} &= \frac{\ii}{2\hbar} \int_0^\infty \dd \tau' \avg{\dot{\hat d}_{0,\mr{st}}(\tau) \dot{\hat d}_{0,\mr{st}}(\tau-\tau')} \avg{\partial_i \hat\phi(\tau) \partial_j \hat\phi(\tau-\tau')} - (\tau \leftrightarrow \tau-\tau') \\
&= \frac{\ii \delta_{ij}}{(2\pi)^26 \hbar c^2} \int_0^\infty \dd \tau' \int_{-\infty}^\infty \dd \omega \int_{-\infty}^\infty \dd \omega' \, \omega^4 \omega'^2 |\alpha_0(\omega)|^2 e^{\ii (\omega + \omega') \tau'} \sbkt{D^+(\omega) D^+(\omega') - D^-(\omega) D^-(\omega')} \\
&= -\frac{\delta_{ij}}{(2\pi)^26 \hbar c^2} \mc{P} \int_{-\infty}^\infty \dd \omega \int_{-\infty}^\infty \dd \omega' \frac{\omega^4 \omega'^2}{\omega + \omega'} |\alpha_0(\omega)|^2 \sbkt{D^+(\omega) D^+(\omega') - D^-(\omega) D^-(\omega')},
\end{align}
in the second line using Eq. \eqref{eq:grad_phi_grad_phi} to simplify the correlator of spatial derivatives. Here \(\mc{P}\) denotes the Cauchy principal value.

\subsection{Time-Dilation Dispersion Potential}
The time-dilation term of the dispersion potential is computed in a similar manner:
\begin{align}
C_{ij}^{(2),\mr{TD}} &= \frac{\ii}{2\hbar} \int_0^\infty \dd \tau \sbkt{\avg{\hat B_i^\mr{TD}(\tau), \hat B_j^\mr{TD}(0)}}\\
&= \frac{\ii a_i a_j}{8 \hbar c^4} \int_0^\infty \dd \tau' \avg{\bkt{\dot{\hat d}_{\mr{st},1}(\tau) - \dot{\hat d}_{\mr{st},0}(\tau)}\bkt{\dot{\hat d}_{\mr{st},1}(\tau-\tau') - \dot{\hat d}_{\mr{st},0}(\tau-\tau')}} \avg{\hat\phi(\tau) \hat\phi(\tau-\tau')} - (\tau \leftrightarrow \tau-\tau') \\
&= \frac{\ii a_i a_j}{(2\pi)^28 \hbar c^4} \int_0^\infty \dd \tau' \int_{-\infty}^\infty \dd \omega \int_{-\infty}^\infty \dd \omega' \, \omega^4 |\alpha_0(\omega) \eta(\omega) - \alpha_0(\omega)|^2  e^{\ii (\omega + \omega') \tau'} \sbkt{D^+(\omega) D^+(\omega') - D^-(\omega) D^-(\omega')} \\
&= -\frac{a_i a_j}{(2\pi)^28 \hbar c^4} \mc{P} \int_{-\infty}^\infty \dd \omega \int_{-\infty}^\infty \dd \omega' \frac{\omega^4}{\omega + \omega'} |\alpha_0(\omega) \eta(\omega) - \alpha_0(\omega)|^2 \sbkt{D^+(\omega) D^+(\omega') - D^-(\omega) D^-(\omega')}.
\end{align}

\end{widetext}

\bibliography{QBM}
\end{document}